\documentstyle[sprocl,epsfig]{article}

\def\Journal#1#2#3#4{{#1} {\bf #2}, #3 (#4)}

\def\NPB{{\em Nucl. Phys.} B}

\def\ZPC{{\em Z. Phys.} C}
\def\PZH{\em Pisma ZhETF}
\def\PR{\em Phys. Rep.}
\def\SJNP{\em Sov. J. Nucl. Phys}

\def\be{\begin{equation}}
\def\ee{\end{equation}}
\def\bq{\begin{eqnarray}}
\def\eq{\end{eqnarray}}

\begin{document}

\title{DEEP INELASTIC SCATTERING AND LIGHT-CONE WAVE FUNCTIONS}

\author{V.M. BELYAEV}

\address{ITEP,  117259  Moscow, Russia.}

\author{Mikkel B. JOHNSON}

\address{LANL, Los Alamos, NM 87545, USA}


\maketitle\abstracts{
In the framework of light-cone QCD sum rules, we study the valence quark
distribution function $q(x_B)$ of a pion for moderate $x_B$.
The sum rule with the leading twist-2 wave function gives
$q(x_B)=\varphi_\pi(x_B)$. Twist-4 wave functions give about 30\% for
$x_B\sim 0.5$. It is shown that QCD sum rule predictions, with the
asymptotic pion wave function, are in good agreement with experimental data.
We found that a two-hump profile for the twist-2 wave function
leads to a valence quark disribution function that contradicts 
 experimental data.
}

\section{Introduction}
Two types of objects in hadron physics are used to
represent quark and gluon distributions in a hadron.
One is the quark and gluon distribution functions that appear in  
deep-inelastic lepton-hadron scattering, and the other is the  light-cone
wave functions of hadrons introduced in pQCD by Chernyak and 
Zhitnitsky \cite{cz} to describe hadron form factors at large $Q^2$.  In this 
paper, we investigate a relationship between these
objects.  To make a prediction for the valence-quark distribution in terms of
light-cone wave functions, we combine the ideas of calculating the 
deep-inelastic scattering amplitudes in the QCD sum rule approach \cite{svz}
as suggested by Ioffe \cite{ioffe} with 
the so-called light-cone QCD sum rules \cite{bal}.

In this work we propose a method to calculate the pion
structure function directly in terms of light-cone wave functions.
The starting point of this method was formulated in Ref.\cite{bi},
where it was pointed that if $x_B=Q^2/(2pq)$ is not close to the
boundary values $x=0$ and $x=1$, then the imaginary part of the deep-inelastic 
scattering amplitude is determined by small distances in the $t$-channel.
A general proof of this statement follows from the fact that
at large $|p^2|$ and $|q^2|$, the nearest singularity in the $t$-channel is 
at $t=-\frac{x_B}{1-x_B}p^2$ in the kinematics that is used in this paper.
Thus, in the case of intermediate $x_B$ we can  use the OPE to
construct QCD sum rules.

\section{Correlator}

To calculate the quark distribution function for a pion, we consider
the following correlator:
\bq
T_{\mu\rho\lambda}(p,q,k)=
-i\int d^4xd^4ze^{ipx+iqz}
<0|T\{ j_\mu^5(x),j_\rho^d(z),j_\lambda^d(0)|\pi^-(k)>,
\label{1}
\eq
where $k$ is a pion momentum, and where
\bq
j_\mu^5=\bar{u}\gamma_\mu\gamma_5d,\;\;\;
j_{\rho}^d=\bar{d}\gamma_\rho d ,
\label{2}
\eq
\bq
k^2=0, \;\;\; q^2=(p+q-k)^2,\;\;\; t=(p-k)^2=0,\;\;\; s=(p+q)^2,
\;\;\;Q^2=-q^2.
\label{3}
\eq
>From relations (\ref{3}) it is easy to determine that
\bq
(2k,p+q)=s+Q^2, \;\;\; (2pk)=p^2 .
\label{4}
\eq

We calculate the discontinuity in $s$ at fixed $p^2$ and $Q^2$ of the
correlator (\ref{1}),
\bq
Im T_{\mu\rho\lambda}=\frac{1}{2i}
\left[
T_{\mu\rho\lambda}(p^2,q^2,s+i\varepsilon)-
T_{\mu\rho\lambda}(p^2,q^2,s-i\varepsilon)
\right] ,
\label{5}
\eq
where $p^2$ and $q^2$ are space-like  vectors, $p^2<0$, $q^2<0$, such
that $|p^2|,|q^2|\gg \Lambda_{QCD}$.
In the scaling limit, we assume that $|p^2|\ll |q^2|$ and keep only the
first nonvanishing terms in an expansion in powers of $p^2/q^2$.
Perturbative logarithmic corrections are taken into account
by the Alterelli-Parisi equation.

We calculate $Im T_{\mu\rho\lambda}$ in the physical region of the 
$s$-channel, and the pion contribution in this amplitude
has the following form:
\bq
Im T_{\mu\rho\lambda}=p_\mu\frac{f_\pi}{p^2}
Im\left\{
\int d^4ze^{iqz}<\pi(p)|T\{ j_\rho^d(z),j_\lambda^d(0)\}|\pi(k)>
\right\} .
\label{6}
\eq
The general form for
$Im\left\{
\int d^4ze^{iqz}<\pi(p)|T\{ j_\rho^d(z),j_\lambda^d(0)\}|\pi(k)>
\right\}$ is
\bq
A_1(s,Q^2)(q^2g_{\rho\lambda}-q_\rho^{(2)} q^{(1)}_\lambda)
\nonumber
\\
+A_2(s,Q^2)(q^2g_{\rho\lambda}
-q^{(1)}_\rho q^{(1)}_\lambda-q^{(2)}_\rho q^{(2)}_\lambda +
q^{(1)}_\rho q^{(2)}_\lambda)
\nonumber
\\
+
B_1(s,Q^2)\left(p-\frac{pq^{(1)}}{q^2}q^{(1)}\right)_\rho
\left(p-\frac{pq^{(2)}}{q^2}q^{(1)}\right)
_\lambda
\nonumber
\\
+B_2(s,Q^2)\left(p-\frac{pq^{(1)}}{q^2}q^{(1)}\right)_\rho
\left(p-\frac{pq^{(2)}}{q^2}q^{(2)}\right)_\lambda 
\nonumber
\\
+B_3(s,Q^2)\left(p-\frac{pq^{(1)}}{q^2}q^{(2)}\right)_\rho
\left(p-\frac{pq^{(2)}}{q^2}q^{(1)}\right)_\lambda
\nonumber
\\
+B_4(s,Q^2)\left(p-\frac{pq^{(1)}}{q^2}q^{(2)}\right)_\rho
\left(p-\frac{pq^{(2)}}{q^2}q^{(2)}\right)_\lambda
\label{7}
\eq
where $q^{(1)}=q$ and $q^{(2)}=p+q-k$ are the momenta of the virtual
photons; $(q^{(1)})^2=(q^{(2)})^2=q^2$.
It is clear that
\bq
4 x_B^2 q^d(x_B) =\frac{Q^2}\pi\left(B_1(s,Q^2)+B_2(s,Q^2)+B_3(s,Q^2)+B_4(s,Q^2)
\right)\;\;\;Q^2\rightarrow\infty
\label{8}
\eq
where
\bq
Im\left\{
\int d^4ze^{iqz}<\pi(p)|T\{ j_\rho^d(z),j_\lambda^d(0)\}|\pi(p)>
\right\}
\nonumber
\\
=
4\pi\frac{x_B^2q^d(x_B)}{Q^2}
\left(p-\frac{pq}{q^2}q\right)_\rho\left(p-\frac{pq}{q^2}q\right)_\lambda
+...~.
\label{9}
\eq
Here $q^d(x_B)$ is $d$-quark distribution function of a pion.

To find the quark distribution function in this paper, wwe consider 
the tensor structure $p_\mu p_\rho p_\lambda$ in
correlator (\ref{1}). We define the imaginary part of the correlation 
function for these tensor structures as $\frac{4 \pi x_B^2}{Q^2}t(p^2,x_B)$.
We can write the following dispersion relation for this function:
\bq
t(p^2,x_B)=f_\pi
\left(\frac{q^d(x_B)}{p^2}+\int ds \frac{\rho(s,x_B)}{s-p^2}ds\right) ,
\label{10}
\eq
where the second term in the right-hand side is the higher-states contribution.
To supress the exited states contribution, as usually done in QCD sum rules,
we will consider instead of $t(p^2,x)$ its Borel transform in $p^2$:
\bq
t(M^2,x_B)=\frac{(-p^2)^{n+1}}{n!}\left(
\frac{d}{dp^2}\right)^n t(p^2,x_B)
\nonumber
\\
=-f_\pi\left(x_B^2 q^d(x_B)+
\int \rho(s,x_B)e^{-s/M^2}ds\right) .
\label{11}
\eq

\section{QCD sum rule}

To find $q^d(x_B)$, we apply the OPE near 
the light-cone $x^2=0$. In contrast to conventional QCD sum rules based on
the OPE of a T-product of currents at small distances, we apply an expansion
near the light-cone expressed in terms of nonlocal operators, i.e. matrix
elements that define hadron light-cone wave functions of increasing twist.
The amplitude of two-quark nonlocal operator has the following form:
\bq
<0|\bar{u}(x)\gamma_\mu\gamma_5d(0)|\pi(k)>=
ik_\mu f_\pi\int_0^1due^{-ikxu}(\varphi_\pi(u)+x^2g_1(u)+O(x^4))
\nonumber
\\
+f_\pi\left(x_\mu-\frac{x^2}{kx}k_\mu\right)
\int_0^1due^{-ikxu}(g_2(u)+O(x^2)) ,
\label{13}
\eq
where $\varphi_\pi(u)$, and $g_1(u)$ and $g_2(u)$, are the 
leading twist-2 and the twist-4 pion light-cone wave functions, respectively.

The leading twist-2 operator gives the following contribution to $t(p^2,x)$
\bq
f_\pi\frac{\varphi_\pi(x_B)}{p^2} .
\label{12}
\eq
Using eq.(\ref{10}) for $t(p^2,x_B)$, the leading twist-2 wave 
function contribution is
\bq
 q^d(x_B)=\varphi_\pi(x_B) .
\label{14}
\eq
It is clear that this relation corresponds to a pure parton picture 
in which the pion consists of two valence quarks  only:
\bq
\int_0^1 q^d(x_B)dx_B=1,\;\;\;\int_0^1x_B q^d(x_B)dx_B=1/2 .
\label{15}
\eq
These relations follow from the normalization of the twist-2 pion wave function
($\int_0^1\varphi_\pi(u)du=1$), and from its symmetry: 
$\varphi_\pi(u)=\varphi_\pi(1-u)$.

In this paper we use the asymptotic pion wave function:
\bq
\varphi_\pi(u)=\varphi^{(asym.)}(u)=6u(1-u) .
\label{16}
\eq
Note that there are alternative models for $\varphi_\pi$:  the
Chernyak-Zhitnitsky wave functions $\varphi_\pi^{CZ}$ and
the function that was introduced in Ref.\cite{br}.  Because of their
two-humped shape, these alternative models for the pion wave function lead to
poor agreement between the QCD sum rule prediction and experiment.

To make a reasonable prediction, we have to estimate the twist-4 
wave function contribution.  In this paper, we take into account the 
contribution of the two-particle wave functions of twist-4.
This contribution to $t(p^2,x_B)$ is
\bq
\frac{f_\pi}{p^4}f_4(x_B)=\frac{4f_\pi}{p^4}\left(
\frac{g_1(x_B)+G_2(x_B)}{x_B}+\frac12g_2(x_B)-\frac{d g_1(x_B)}{dx_B}
\right). 
\label{17}
\eq
where $G_2(u)=-\int_0^u dv g_2(v)$.

Using eqs.(\ref{10},\ref{12},\ref{16}), after making a Borel transformation,
we have
\bq
\varphi_\pi(x_B)-\frac{4}{ M^2}\left(
\frac{g_1(x_B)+G_2(x_B)}{x}+\frac12g_2(x_B)-\frac{d g_1(x_B)}{x_B}
\right)
\nonumber
\\
=q^d(x_B)+C(x_B)e^{-m_{A_1}^2/M^2} .
\label{18}
\eq
Here, we take into account the contribution of the $A_1$-meson to evaluate
the higher states contribution.

To estimate twist-4 contribution, we use the following set of
twist-4 pion wave functions (see Ref.\cite{br} and Ref.\cite{g}):
\bq
g_1(u)=\delta^2\frac52+\frac12\varepsilon\delta^2
\left(
\bar{u}u(2+13\bar{u}u\right.
\nonumber
\\
\left.
+2u^3(10-15u+6 u^2)\ln(u)
+2\bar{u}^3(10-15\bar{u}+6\bar{u}^2)\ln(\bar{u})\right) ,
\nonumber
\\
g_2(u)=\frac{10}3\delta^2\bar{u}u(u-\bar{u}) ,
\label{19}
\eq
where $\bar{u}\equiv1-u$.  One of the parameters is defined by the matrix 
element
\bq
<0|g_s\bar{u}\tilde{G}_{\alpha\beta}\gamma_\beta u|\pi(p)>=
i\delta^2f_\pi p_\mu .
\label{20}
\eq
The QCD sum rule estimate of Ref.\cite{nsvz} yelds $\delta^2=0.2GeV^2$ at
$\mu\simeq 1 GeV$.  At $\mu\simeq 1GeV$, $\varepsilon\simeq 0.5$
(see Ref.\cite{br}).

To fix the function $C(x_B)$, we take the QCD sum rule
(\ref{17}) in the limit $M^2\rightarrow\infty$ to be valid up to
terms $1/M^2$. This condition leads to the relation:
$
C(x_B)=\frac{f_4(x_B)}{m_{A_1}^2}$.

Finally we obtain the following QCD sum rule:
\bq
q^d(x_B)=\varphi_\pi(x_B)-f_4(x_B)\left(\frac1{M^2}+\frac1{m_{A_1}^2}
e^{-m_{A_1}^2/M^2}\right) .
\label{21}
\eq

For $M^2\sim 1GeV^2$ and $0.2<x_B<0.8$, the contribution of $A_1$-meson,
which imitates the higher state contribution, and the contribution
of twist-4 wave functions, is less than 30\%. The results are
depicted in Fig.1. 

Note that in our calculations we do not take into consideration
the contribution of leading logarithms. So, in order to make a comparison
with experiment, we have to use the quark distribution function at low
$Q^2$ where these logarithms are small. 
Evolution of experimental data to low $Q^2$ has been done in 
Ref.\cite{vogt}, and we use the results of this
paper as experimental data.

\begin{figure}
\begin{center}
\vspace*{-0.3cm}
\epsfig{file=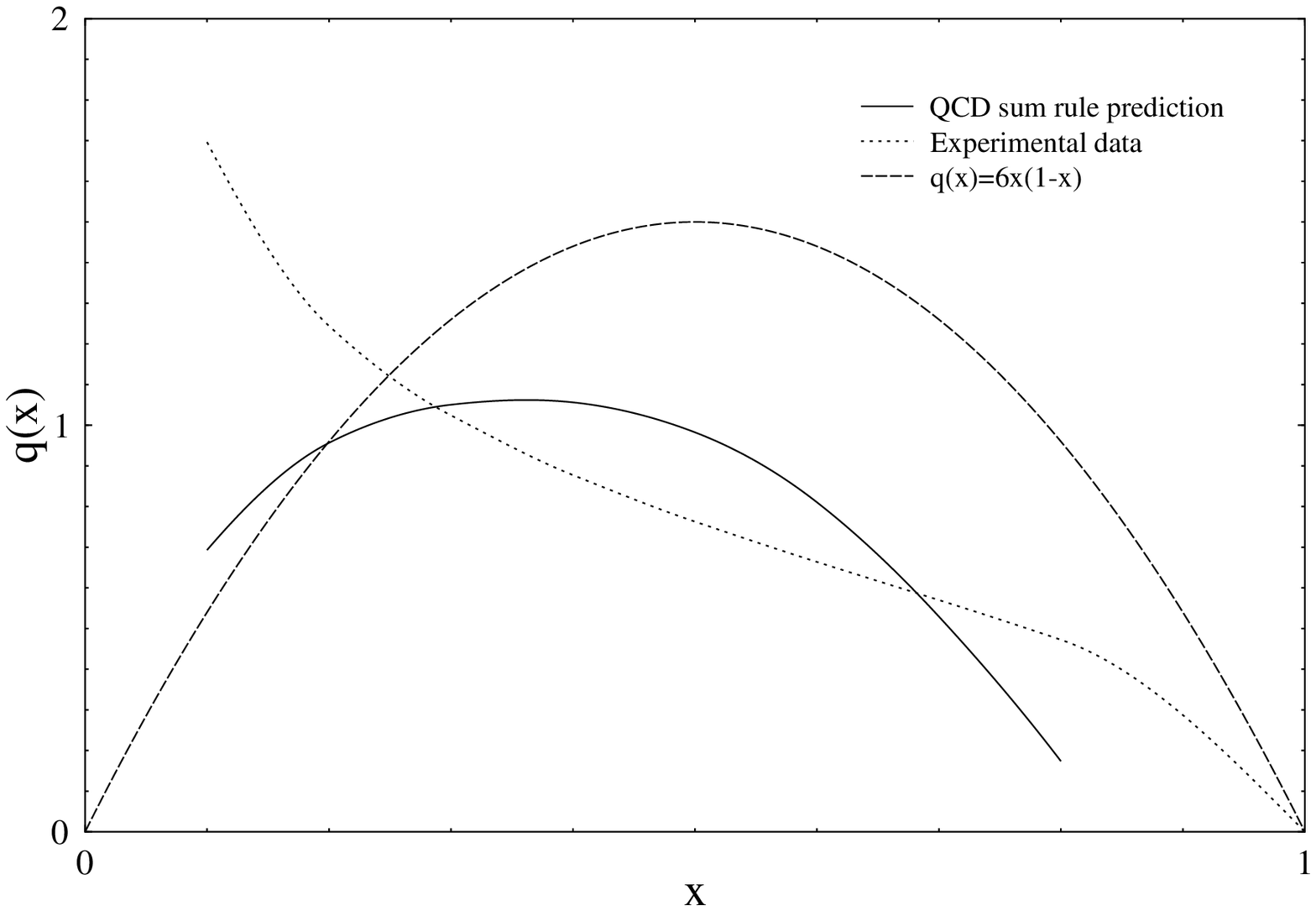,height=7.0cm}
\vspace*{-0.9cm}
\end{center}
\vspace*{-2.5cm}
\caption[]
 {}
\end{figure} 

It is interesting to estimate the second momemnt of the quark distribution
function. Assuming that the region near the end points $x_B=0,1$ 
(where our consideration is not valid) gives
a small contribution to the second moment, the sum rule (\ref{21})
gives the following sum rule for the second moment:
\bq
\frac12-\frac{\delta^2}{M^2}=M_2^d+(  higher\;\;resonance\;\; contribution).
\label{22}
\eq
The analysis of this sum rule gives $M_2^d=0.27\pm 0.05$ at low
normalization point, which is in a good agreement with experimental data:
$M_2^d\simeq 0.3$ (see Ref.\cite{vogt}).

\section{Conclusions}

The light-cone QCD sum rules considered in  this paper lead to satisfactory 
agreement with experimental data.  To improve the prediction of the QCD 
sum rule, it is nessesary to take into consideration twist-4 quark-gluon 
wave functions.  One should note that the present experimental data for the 
quark distribution functions was obtained from an analysis of the Drell-Yan 
process, which is uncertain through the so-called {\bf K}-factor.
The choice $\varphi_\pi=\varphi^{(asym.)}$ gives the best agreement
between the QCD sum rule prediction and the experimental data.
Alternative models for the pion wave function having a two-humped profile
lead to a poor agreement with the experimental data.

\section{Acknowledgements}  This research was sponsored by the U.S. Department
of Energy at Los Alamos National Laboratory under contract W-7405-ENG-3.


\section*{References}
\begin{thebibliography}{99}

\bibitem{cz}V.L. Chernyak and A.R. Zhitnitsky, \Journal{\PR}{112}{173}{1984}.

\bibitem{svz}M.A. Shifman, A.I. Vainshtein and V.I. Zakharov,
\Journal{NPB}{147}{285}{1979}; B {\bf 147},{448} {(1979)}.

\bibitem{ioffe}B.L. Ioffe, \Journal{\PZH}{42}{266}{1985};
{\bf 43}, {316} {(1986)}.

\bibitem{bi}V.M. Belyaev and B.L. Ioffe, \Journal{\NPB}{310}{548}{1988}. 

\bibitem{bal}I.I. Balitsky, V.M. Braun, and A.V. Kolesnichenko,
\Journal{\SJNP}{44}{1028}{1986}; \Journal{\NPB}{312}{509}{1989}.


\bibitem{br}V.M. Braun, and I.B. Filyanov, \Journal{ZPC}{48}{239}{(1990)}.


\bibitem{g}A.S. Gorsky,\Journal{\SJNP}{41}{1008}{1985};
{\bf 45}, 212 (1987); {\bf 50}, 498 (1989).

\bibitem{nsvz}V.A. Novikov, M.A. Shifman, A.I. Vainshtein and V.I. Zakharov,
\Journal{\NPB}{237}{525}{1984}.

\bibitem{vogt}M. Gl\"uck, E. Reya, A. Vogt,
\Journal{\ZPC}{53}{651}{1992}. 

\end{thebibliography}
\end{document}